\documentclass[prl,showpacs,twocolumn,floatfix,a4paper,twoside]{revtex4}
\usepackage{graphics, subfigure, epsfig, color, moreverb, eufrak}
\graphicspath{{./figs/}}
\newcommand{\ddt}{\frac{\partial}{\partial t}}
\newcommand{\ihddt}{i\hbar\ddt}
\newcommand{\qav}[1]{ \langle #1 \rangle} 
\newcommand{\pa}[1]{\left ( #1 \right )}                
\newcommand{\ab}[1]{\left | #1 \right |}                
\newcommand{\eqa}[1]{\begin{eqnarray}#1 \end{eqnarray}} 
\newcommand{\bi}{\begin{itemize}}
\newcommand{\ei}{\end{itemize}}
\newcommand{\be}{\begin{enumerate}}
\newcommand{\ee}{\end{enumerate}}
\newcommand{\reff}[1]{Fig.~\ref{#1}}
\newcommand{\refe}[1]{Eq.~(\ref{#1})}

\renewcommand{\vec}[1]{\mathbf{#1}}

\newcommand{\mrm}[1]{\mathrm{#1}}

\definecolor{light-gray}{gray}{.85}
\begin{document}
\title{Characterization of Disorder in Semiconductors via Single--Photon Interferometry\\
}
\author{P. Bozsoki$^{1}$, P. Thomas$^1$, M. Kira$^1$, W. Hoyer$^1$, T.~Meier$^1$, S.W. Koch$^1$, K. Maschke$^2$, I. Varga$^3$, H.~Stolz$^4$}
\affiliation{$^1$Department of Physics and Material Sciences Center,
Philipps-Universit\"at Marburg, Germany\\
$^2$Institut de Th\'eorie des Ph\'enom\`enes Physiques,
Ecole Polytechnique F\'ed\'erale, CH-1015 Lausanne, Switzerland\\
$^3$Elm\'eleti Fizika Tansz\'ek,
Fizikai Int\'ezet, Budapesti M\H{u}szaki \'es
Gazdas\'agtudom\'anyi Egyetem, Hungary\\
$^4$ Institut f\"ur Physik, Universit\"at Rostock, Universit\"atsplatz 3,
D-18051 Rostock, Germany}

\date{\today}

\begin{abstract}
The method of angular photonic correlations of spontaneous
emission is introduced as an experimental, purely optical
scheme to characterize disorder in semiconductor
nanostructures. The theoretical expression for the angular correlations is
derived and numerically evaluated for a model system.
The results demonstrate how the proposed experimental
method yields direct information
about the spatial distribution of the relevant states and thus on the
disorder present in the system.
\end{abstract}


\maketitle

Disorder in solids is a topic of increasing interest in solid-state
research in the last decades.  Intriguing and widely celebrated phenomena
likeQuantum-Hall-Effect, photon echoes, and Anderson localization rely on
the existence of disorder. However, the significance of disorder is not
purely academic. Modern devices are fabricated on smaller and smaller
length scales. Systems  with reduced dimensionality can be realized by
semiconductor  heterostructures.  Concomitantly, disorder due to interface
roughness and/or compositional fluctuations plays a crucial role in the
performance of these nanostructures. Clearly, for a detailed  understanding
of the influence of disorder on the investigated phenomena, it is
important to know to what degree the electronic states deviate from  the
Bloch-like character which governs the electronic properties in  perfectly
periodic structures. In particular, localized electronic  states can occur
which strongly alter both the optical and transport  properties of a given
system~\cite{ot, es, ei, mtk, Yayon:PRL,hardToMapDisorderToOptics-1,
hardToMapDisorderToOptics-2}. It is, therefore, important to have a
reliable method that allows the experimentalist to quantitatively
characterize the disorder in a given structure without its destruction.
Unfortunately, in this respect the present situation is not very
satisfactory.

For example, the lattice disorder in alloy semiconductors can be determined in
principle by spatially resolved atomic--level microscopies~\cite{nature,
nature-afm, nature-stm, nature-stm2, nature-tem}, but its influence on
e.g.\ optical properties is still not well
understood~\cite{Yayon:PRL,hardToMapDisorderToOptics-1,
hardToMapDisorderToOptics-2}.
Alternatively, one can obtain information about the connection of disorder
and optical properties directly within the resolution limit set by the
wavelength of light. Examples include micro-(nano)-luminescence
\cite{wegener-1, wegener-2}, near-field optical imaging \cite{nr4-1,nr4-2},
 and resonant Rayleigh  scattering (RRS)
\cite{schwedt2006,resram,Langbein:PRL2002}.  In order to deduce the spatial
dependence of the disorder potential, the first two methods require a
spatial scanning of the sample.  When the angular dependence of the RRS
intensity is measured, one can determine the energy statistics of
emitters~\cite{resram}, as well as the localization length scales of the
optically active excitonic
states~\cite{schwedt2006,resram,Langbein:PRL2002}. However, RRS can deduce
the exact disorder potential and position of emitters only indirectly via
an inverse-scattering problem.

In this Letter, we suggest a quantum--optical interferometric method which
provides the disorder potential directly. Our scheme is based on the
photonic correlations that are determined from the stationary spontaneous
emission by collecting two different emission directions into a common
detector. Recent experiments have shown that strong single--photon
interferences can be observed in this setup~\cite{walter}. We demonstrate
that the qualitative features of these interferences can be utilized to
reconstruct the disorder landscape of the light emitters for length scales
varying between the used wavelength and the illumination spot size. We
also show that the reconstruction of disorder potential can be performed
solely on the basis of experimental data and that the method is robust to
experimental inaccuracies. In contrast with
micro-(nano)-luminescence~\cite{wegener-1,wegener-2} and near-field optical
imaging~\cite{nr4-1,nr4-2}, our scheme does not need spatial scanning.

To present the idea behind our proposed
quantum--optical method, we need to quantize the light
field and compute the correlations among the emitted photons including
the disorder as well as the relevant many--body
interactions among the electron--hole excitations in the
semiconductor heterostructure. For this purpose, and in order to
keep our analysis as transparent as possible, we choose a simple
model based on a one--dimensional tight--binding description. For
a chain of $N$ sites with periodic boundary conditions, every site, if
isolated from the others, supports one electron and one hole state with the
energies $\epsilon_j^e$ and $\epsilon_j^h$, respectively. The index
$j$ refers to the position of the site and the energies $\epsilon_j^{e,h}$ vary
from site-to-site due to disorder. The corresponding carrier system is
described by the material Hamiltonian
\eqa{
{H}_\mrm{mat}&=& \sum_j\epsilon_j^e  e^{\dag}_j e_j
+ J^e\sum_{<jl>} e^{\dag}_j  e_l
+ \sum_j\epsilon_j^h  h^{\dag}_j  h_j\nonumber\\
&+& J^h\sum_{<jl>}  h^{\dag}_j  h_l - \sum_{j,l} V_{jl}\,
e^{\dag}_{j}  h^{\dag}_{l} h_{l}  e_{j}
\label{eq:Hmatter}
}
where $ e^{\dag}_{j},  e_{j}$ and $ h^{\dag}_{j},  h_{j}$ denote the
electron and hole creation and annihilation operators  at a site $j$. The
constant $J^{e,h}$ defines the bandstructure via the nearest neighbor
coupling indicated by the  symbol $<jl>$. We concentrate here on
low--density conditions such that we only describe the attractive Coulomb
interaction of electron--hole pairs, given by the last term in
\refe{eq:Hmatter}. The model parameters are chosen to reproduce the
effective mass and the exciton binding energy of typical GaAs quantum--well
systems~\cite{mtk}. In particular, $N$ sites are positioned evenly at
distances $a$ with $J^e=-5.737$~meV and $J^h=0.835$~meV. The
regularized Coulomb matrix element is given by $V_{jl} = U_0/(|j-l|a +
a_0)$ with $U_0=7$~meV and $a_0=0.5\,a$. This yields an exciton binding
energy of $E_b=7.5$~meV.

The quantized light field can be expressed via a mode expansion in terms of
plane--wave modes $u_\vec{q}{\pa{\vec{r}}} = e^{i\vec{q}\cdot\vec{r}}$ with
momenta $\vec{q}$~\cite{cohen,PQE} and photon operators $B_\vec{q}$. The
light--matter interaction is given by ${H}_\mrm{L}={\sum_{\vec
q}\hbar\omega_q (B_{\vec q}^{\dag}B_{\vec q}+1/2)} - \sum_{\vec{q}, j}
({i\eta u_{\vec{q},j} B_\vec{q} e^\dag_j h^\dag_j + \mrm{h.c.}})$, where
the absorption of a photon creates an electron--hole pair at the site $j$
since only direct transitions are allowed. This term is proportional to
$\eta=\mu_0 {\mathcal E}$ with the dipole--matrix element $\mu_0$ and the
vacuum--field amplitude ${\mathcal E}$. The light--matter interaction also
depends on the mode function $u_{\vec{q},j}$ at the position of the site,
$\vec{R}_j$. The sites can be assummed to be equally spaced since their
size is much smaller than the optical wavelength. Consequently, we may
choose $\vec{R}_j = R_j \vec{e}_{\|} = aj \vec{e}_{\|}$ such that all
disorder effects enter via the energies $\epsilon^{e,h}_j$. The
corresponding $u_{\vec{q},j} = e^{i R_j q_{\|}}$ contains the in-plane
photon momentum in the direction $\vec{e}_{\|}$.

Under incoherent conditions, all coherent quantities, $\qav{h_j e_j}$,
$\qav{e_j^\dag h_j^\dag} $ and $\qav{B_\vec{q}}$ , $\qav{B_\vec{q}^\dag}$,
vanish. In this situation, the system emits light only spontaneously
leading to non--vanishing $\qav{{B_\vec{q}}^\dag B_\vec{q'}}$. For
steady--state conditions, the photon--flux $\partial/{\partial t}
\qav{B_\vec{q}^\dag B_\vec{q}}$ defines the photoluminescence (PL)
intensity while $\partial/{\partial t} \qav{B_\vec{q}^\dag B_{\vec{q}'}}$
determines the quantum-optical correlations between two different emission
directions. By using a suitable optical arrangement, one can collect light
in the directions $\vec{q}$ and  $\vec{q}' \ne \vec{q}$ to a common detector.
With this interferometric setup, $U_{\vec{q}, \vec{q}'} \equiv
\partial/{\partial t} \qav{B_\vec{q}^\dag B_{\vec{q}'}}$ defines the
contrast of the single--photon interferences \cite{walter} for
steady--state emission conditions. This quantity follows from
\begin{equation} \label{eq:flux}
\hbar U_{\vec{q}, \vec{q}'} =
-  \sum_l  \pa{{\eta^* u_{\vec{q'},l}^*}{ } \Pi^\vec{q}_{l,l}
+ {\eta  u_{\vec{q},l}}
{\Pi^\vec{q'}_{l,l}}^*}\,,
\end{equation}
showing that $U_{\vec{q}, \vec{q}'}$ is coupled to the photon--assisted
polarization $\Pi^\vec{q}_{j,l} \equiv \qav {B^\dag_\vec{q} h_j e_l}$ and
its dynamics
\begin{eqnarray} \label{eq:pum}
&&\ihddt \Pi^\vec{q}_{j,l} =
 \pa{\epsilon_{j,l} - \hbar \omega_q} \Pi^\vec{q}_{j,l}
 +   J^e  \sum_{\Delta=\pm 1} \Pi^\vec{q}_{j,l+\Delta} \nonumber\\
&& + J^h \sum_{\Delta=\pm 1} \Pi^\vec{q}_{j+\Delta,l}
 + \ihddt \Pi_{j,l}^\vec{q} |^\mrm{scatt} + \ihddt \Pi_{j,l}^\vec{q} |^\mrm{stim}
\nonumber\\
&& -  V_{lj} \pa{1-f_j^h - f_l^e} \Pi^\vec{q}_{j,l}
+ i \eta u_{\vec{q},j} S_{jl} \,\, ,
\end{eqnarray}
where $\epsilon_{j,l} = \epsilon_l^{e}+\epsilon_j^{h}-\sum_m \pa{  V_{lm}
f_m^{e}+  V_{jm}  f_m^{h}}$. The second line in Eq.~(3) contains
contributions due to Coulomb scattering and stimulated emission. The
stimulated term can be omitted for systems without a cavity~\cite{PQE}. As
another simplification, we introduce a constant dephasing $\hbar \ddt
\Pi_{j,l}^\vec{q} |^\mrm{scatt}  = - \gamma \Pi_{j,l}^\vec{q}$. When the
carrier system is in a quasi-equilibrium, $S_{jl}=\sum_m u_{\vec{q},m}
/u_{\vec{q},j} \langle e_m^{\dag}h_m^{\dag}h_j e_l\rangle$ acts as a
constant spontaneous--emission source and the Coulomb term, $V\Pi$, leads
to excitonic resonances in the emission. At the lowest level, $S_{jl}$ is
given by its Hartree--Fock approximation $S_{jl}= \delta_{jl} f_j^e f_j^h$
containing the excited state occupations, $f_j^e=\qav{e_j^\dag e_j}$ and
$f_j^h=\qav{h_j^\dag h_j}$.

In order to understand how disorder effects enter in
$U_{\vec{q},\vec{q}'}$, we first solve Eqs.~(\ref{eq:flux})--(\ref{eq:pum})
for a simplified case by setting all carrier--interaction terms to zero
$J^{e,h}_j=V_{jl}=0$. This yields a steady--state expression
\begin{equation} \label{eq:U-TLS}
 U_{\hbar\omega}(\Delta q) \equiv U_{{\mathbf q},{\mathbf q'}}=
{\mathcal N} \sum_j \frac{\gamma S_{jj} e^{i\Delta q
R_j}}{\pa{\epsilon_j-\hbar\omega_q}^2
+ \gamma^2} \,\,,
\end{equation}
where $\epsilon_j=\epsilon^{e}_j + \epsilon^{h}_j$. We have also assumed
that $\ab{\vec{q}} = \ab{\vec{q}'}$ differ only by the emission direction
giving $\omega=c\ab{\vec{q}}$ and $\Delta q = q_{\|} - q_{\|}'$. The
normalization $\mathcal N$ is chosen such that the correct $100 \%$
interference contrast is observed for $\vec{q} = \vec{q}'$. This form
clearly shows that even the spontaneous emission contains a phase,
$e^{i\Delta q R_j}$, influenced by the position of the emitters. This phase
can always be observed as single--photon interferences~\cite{walls}, which
is the fundamental principle behind the proposed interferometry.

In the continuum limit $a\rightarrow 0$, \refe{eq:U-TLS} reduces to
\begin{equation}\label{eq:corrFuncTLS}
U_{\hbar\omega} \pa{\Delta q} = {\mathcal N}
\int \frac{\gamma S(x)}{\pa{\epsilon(x)-\hbar\omega  }^2 + \gamma^2}
e^{i\Delta q x}dx \,,
\label{corrfct}
\end{equation}
where $S(x)$ defines the level of excitation. We now notice that Fourier
transformation of \refe{eq:corrFuncTLS} with respect to $\Delta q$ produces
a peak at $\epsilon(x)=\hbar \omega$; i.e.\ the position of the emitters
can be determined. Thus we define
\begin{equation}\label{eq:EmitterPos}
U_{\hbar\omega}(x) = \int_{-\Delta q_0}^{\Delta q_0}
U_{\hbar\omega}\pa{\Delta q} e^{-i \Delta q x} d\Delta q ,
\label{pum2}
\end{equation}
where $\Delta q_0=2\omega/c=4\pi/\lambda_0$ is the largest in-plane
momentum which is accessible in principle. This allows us to reconstruct
the disorder potential up to the limits set by $\Delta q_0$ via
\begin{equation} \label{eq:potReconstruction}
U(x) = {\int_{E_1}^{E_2}
\hbar\omega \ab{U_{\hbar\omega}(x)} d\hbar\omega }/
{\int_{E_1}^{E_2}
\ab{U_{\hbar\omega}(x) } d\hbar\omega } \, .
\end{equation}
The integration limits are set by the relevant spectral range
$E_1<\hbar\omega<E_2$.
Equations~(\ref{eq:EmitterPos})--(\ref{eq:potReconstruction}) suggest a
general interferometric scheme to determine the disorder landscape. First, the
interference contrast $U_{\hbar\omega}\pa{\Delta {q}}$ has to be measured
as function of energy for different emission directions. Second, a simple
Fourier transform of $U_{\hbar\omega}\pa{\Delta {q}}$ produces directly the
disorder landscape of emitters.
\begin{figure}[t]
\centering
\includegraphics[width=60 mm, angle=-90]{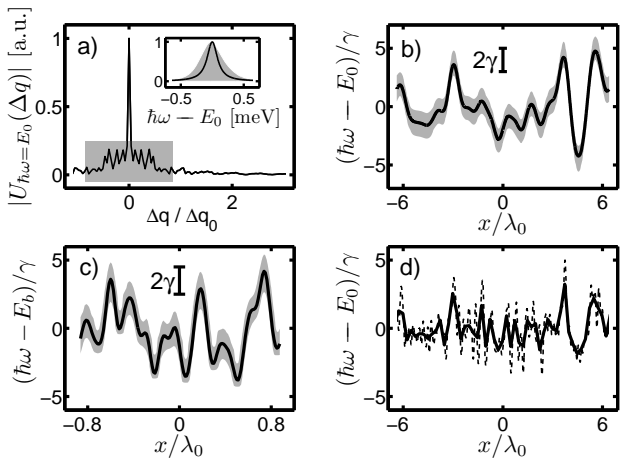}
\caption{a) Correlation function $U_{\hbar\omega=E_0}(\Delta q)$ as a
function of $\Delta q/\Delta q_0$ for noninteracting TLS. The inset shows
PL for the ordered (line) and the disordered (shaded) case. b) Contour plot
of the spatially dependent correlation function $U_{\hbar\omega}(x)$. Solid
line is the input potential with ${\cal L}=1.2\lambda_0$. c) The same for
the fully interacting system,  ${\cal L}=0.2\lambda_0$. $E_b$ is the sum of
$E_0$ and the binding energy. d) Dashed line: input potential with ${\cal
L}=0.25\lambda_0$, based on the same initial distribution of site energies
for noninteracting TLS as b). Solid line: $U(x)$, c.p. with solid line in
b).}
\label{fig:1}
\end{figure}

In our full numerical computations, we generate the energies $\epsilon_j^e$
and $\epsilon_j^h$ from a random box--like distribution  of width $W^e$ and
$W^h$, respectively. We take $W^e/W^h=|J^e/J^h|$, which guarantees that the
disorder amplitudes scale with the effective masses. The ordered case is
recovered by setting $W^e=W^h=0$, i.e., $\epsilon_j^{e,h}=\epsilon^{e,h}$.
The length scale of the disorder potential $\cal L$ is defined by a
Gaussian low--pass filter applied in Fourier space with a maximum spatial
frequency $\propto {\cal L}^{-1}$. We use $\lambda_0=800$~nm, band gap
energy $E_0=1.5$~eV, $\gamma = 0.1$~meV, $N=2048$, and $a=5$~nm.

To illustrate the single--photon interferometric scheme, we first consider
the non-interacting case and set the excitation $S(x)$ to a constant value
within a 10~$\mu$m excitation spot. The inset in \reff{fig:1}a shows the
photoluminescence for $W^e=0.5$~meV (shaded area) and the ordered case
(solid line); clear inhomogeneous broadening is observed. The interference
contrast, $U_{\hbar\omega}\pa{\Delta q}$ is presented in \reff{fig:1}a for
$\hbar\omega=E_0$ as function of $\Delta q$. The central peak at $\Delta q
= 0$ produces a 100\% interference constrast while $U$ drops quickly for
elevated $\ab{\Delta q}$. The peak structure at small $\ab{\Delta q}$
values contains direct information about the disorder. The width of both
central and side peaks is $2\pi/\cal{S}$ where $\cal{S}$ is the size of the
spot. After $U_{\hbar\omega} \pa{\Delta q}$ is defined for all relevant
frequencies, we may construct $U_{\hbar\omega}(x)$ via
\refe{eq:EmitterPos}. We use the maximum optically allowed $\Delta q$
(denoted by shaded area in \reff{fig:1}a) as the integration limits. The
generated $U_{\hbar\omega}\pa{x} \ge \frac12$ contour is shown in
\reff{fig:1}b as shaded area, denoting the full width of half maximum of
$U_{\hbar\omega}\pa{x}$, while the original potential is given by the solid
line. Clearly the center of $U_{\hbar\omega}(x)$ reproduces excellently the
$x$ dependence of the actual disorder landscape. For a fixed $x$,
$U_{\hbar\omega}(x)$ gives \emph{the homogeneous line  for the $\mu$PL
spectrum of the emitter at position $x$.} In the present paper, this width 
is determined by the constant $\gamma$. In reality, microscopic carrier 
and phonon scattering can lead to a position dependent homogeneous line 
width which should be observable in the same way. The total inhomogeneous 
spectrum follows as an integral over $x$. In \reff{fig:1}d we show results 
for a disorder potential with ${\cal L} = 0.25\lambda_0$. The long-range 
features of this potential are very well reproduced by our method.

From an experimental point of view it is critical to know i) the range of
$\vec{q}$-vectors which can be detected and ii) how accurately the side peaks of $U$
must be measured with respect to the central peak. Concerning i), the experimental setup from Ref.~\cite{schwedt2006} would allow for
a range of $\Delta q = N.A./n_{GaAs}\simeq 0.12 \Delta q_0$ only ({\it N.A.}\ is the numerical aperture).
Augmenting the setup by a solid state immersion lens with high index of
refraction (for a discussion see \cite{barnes2002}), and using two optimal
positioned microscope objectives $\Delta q \simeq \pm 0.85 \Delta q_0$ can
be achieved. This range is indicated by the shaded area in \reff{fig:1}a.
Concerning ii), we have solved \refe{eq:corrFuncTLS} analytically for a sinusoidal
$\epsilon(x)$. In particular, we find that the second peak of
$U_{\hbar\omega}\pa{\Delta q}$ is located at $\Delta q = 4\pi/ \cal{L}$ and
the ratio $r$ between the central peak and the second one is
$r_\mrm{sin}=(\sqrt{1+(\gamma/{\Delta}W)^2}+\gamma/{\Delta}W)^{-2}$, where
${\Delta}W$ is the standard deviation of the potential.
Taking two sinusoidal functions with equal amplitude but different spatial
frequency yields  values for $r$ situated clearly below $r_\mrm{sin}$. For various random potentials, we find $r$ is around $r_\mrm{sin}/2$.
This analysis shows that the ratio $\gamma/{\Delta}W$
defines the accuracy requirements in an experiment. For example, if one
wants to resolve disorder up to ${\Delta}W>\gamma/2$, one has to measure
the interference contrast with 10\% accuracy.


To verify the general validity of the scheme defined by Eqs.
(\ref{eq:EmitterPos})--(\ref{eq:potReconstruction}), we next investigate
the fully interacting case by including the band--structure, disorder, and
Coulomb interaction terms in \refe{eq:pum}. We solve the eigenstates
$\Phi_{jl}^\lambda$ and eigenvalues $\epsilon_\lambda$ of the homogeneous
part of \refe{eq:pum} for steady--state conditions. This procedure
yields~\cite{peter1}
\begin{equation}\label{eq:full}
U_{\hbar\omega}\pa{\Delta q}= \sum_{\lambda,jl}
\frac{ \gamma {\mathcal N}
 \Phi_{jj}^{\lambda}  \Phi_{ll}^{\lambda} e^{i {\Delta q} \pa {R_j + R_l}/2}
}{\pa{E_{\lambda} - \hbar\omega}^2 + \gamma^2} \,\,,
 \end{equation}
for constant $S$. We evaluate \refe{eq:full} numerically for a disorder
realization with $W=5\gamma$. Since our main purpose is to show that the
interaction contributions do not affect the reconstruction procedure
(\ref{eq:EmitterPos})--(\ref{eq:potReconstruction}), we can restrict the
calculations to a rather small system ($N=140$, $a=10$~nm) to obtain numerically
feasible computations. Physically, this implies a small ${\cal
S}=1.4\,\mu$m spotsize. Numerically, this leads to a relatively large discretization in $\Delta q$ which forces us to use the $\Delta q$ data taken from
the entire Brillouin zone in the reconstruction procedure.
Figure~\ref{fig:1}c shows the obtained full $U_{\hbar\omega}\pa{x} \ge
\frac12$ in a  contour plot (shaded area) and compares it with the actual
potential (solid line) \cite{note}. Again, the center of  $U_{\hbar\omega}(x)$ matches well with the original disorder potential, which demonstrates the general
applicability and the robustness of the proposed scheme.

\begin{figure}[t]
\includegraphics[width=5cm,angle=-90]{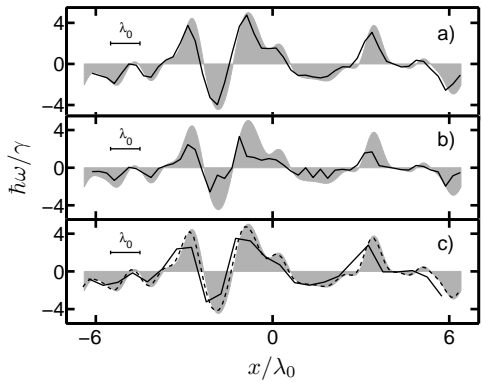}
\caption{Sensitivity of the reconstructed potential to various external
parameters: a) sensitivity of measurement of the interference contrast, b)
random noise, and c) range of momentum scan. Details are in the text. }
\label{fig:4}
\end{figure}

To determine the sensitivity of the reconstruction procedure on several
experimentally relevant inaccuracies, we apply \refe{eq:corrFuncTLS} to
reconstruct the potential via \refe{eq:potReconstruction} for different
situations of the non--interacting case. In all frames of \reff{fig:4} the
shaded area indicates the original potential with $W=5 \gamma$ and the
reconstructed potential $U(x)$ is shown as a solid line. In \reff{fig:4}a,
$U_{\hbar\omega}(x)$ is constructed by omitting values of
$U_{\hbar\omega}\pa{\Delta q}$ with interference contrast below 10\%. In
\reff{fig:4}b, $U_{\hbar\omega}(x)$ is constructed from
$U_{\hbar\omega}\pa{\Delta q}$ where we simply added a 5\% random noise.
When the momentum scan is limited to $\ab{\Delta q} < \Delta q_0/2$ and
$\ab{\Delta q} <2\Delta q_0$ (dashed line) we obtain the $U_{\hbar\omega}(x)$
results presented in \reff{fig:4}c.
This is also accompanied by a limitation of the spatial resolution of the
resulting reconstructed potential. For a minimum $\Delta q_\mrm{min}$
given by experimental limitations, the spot size should be chosen
accordingly, such that $S=2\pi/\Delta q_\mrm{min}$ in order to avoid aliasing
problems.

All these cases indicate that the
proposed photon--correlation interferometry produces the disorder potential
even when a reasonable amount of experimental inaccuracies are present.

In conclusion, we suggest that the experimental method of single--photon
interferometry can directly characterize the optically relevant disorder
landscape in semiconductor heterostructures. The method is only limited by
the wavelength of light and the optical spotsize. We have also demonstrated
a remarkable quality of the reconstructed potential even for weak disorder
and added random noise.

\begin{acknowledgments}
The authors are thankful to H.M. Gibbs, G. Khitrova, R. Zimmermann, and
W. Stolz for valuable discussions. I.V. thanks for financial support from
OTKA under contract 42981 and 46303, K.M. for support from the SNF under
grant 200020-107428. This work has been supported by the Optodynamics
Center of the Philipps-University Marburg, and by the Deutsche
Forschungsgemeinschaft through the Quantum Optics in Semiconductors
Research Group and John von Neumann Institut für Computing (NIC),
Forschungszentrum J{\"u}lich, Germany. T.M. thanks the DFG for support via
a Heisenberg fellowship (ME 1916/1).
\end{acknowledgments}


\begin{thebibliography}{99}
\bibitem{note}{The reconstruction of features smaller than $\lambda_0$ is a direct consequence of our integration over the whole Brillouin zone and will not be observed in an experiment.}
\end{thebibliography}
\end{document}